\documentclass[10pt,a4paper,showpacs,english,aps,pra,twocolumn]{revtex4}
\usepackage{amsmath}
\usepackage{graphicx}
\usepackage{amsfonts}

\makeatletter

\begin{document}

\title{Polaron-like effects in a one-dimensional optical lattice}

\author{M.J. Leskinen, O.H.T. Nummi, F. Massel and P. T\"orm\"a}
\email{paivi.torma@tkk.fi}
\affiliation{Department of Applied Physics, P.O. Box 5100, 02015 Aalto University, Finland}

%\pacs{03.75.Ss, 32.80.-t, 73.50.Lw}

\begin{abstract}
We study a highly imbalanced Fermi gas in a one-dimensional optical lattice from the polaronic point of view. The 
time-evolving block decimationg algorithm
is used to calculate the ground state and dynamics of the system. We find qualitatively similar polaronic behaviour as in 
the recent experiment by Schirotzek et al. \cite{Schirotzek2009a} where radio-frequency spectroscopy was used to observe 
polarons in three-dimensional space. In the weakly interacting limit our exact results are in excellent agreement with a polaron ansatz, 
and in the strongly interacting limit the results match with an approximative solution of the Bethe ansatz,
suggesting a crossover from a quasiparticle to a charge-density excitation regime.
\end{abstract}

\maketitle
\section{Introduction}
Impurity problems are essential in determining low-temperature properties of condensed matter systems. A well-known example is an
electron moving in a crystal lattice and interacting with its surrounding ions \cite{Landau1933a}. This creates lattice polarization and deformation which 
is carried with the electron through the material. Interactions with the ions create an effective potential for the electron and try
to slow it down, which can be modelled as an effective mass for the electron. Another famous impurity problem is the Kondo effect 
\cite{Kondo1964a} where
the scatterings of conduction electrons with magnetic impurities give rise to electrical resistivity. Ultracoldic atom gases provide an excellent playground
for the study of impurity problems due to the controllability of the system parameters. 
For instance, two-component gases can be realized by using two different hyperfine spin states of alkali atoms.
The population imbalance can be controlled by transferring particles between the two hyperfine states with radio-frequency pulses. The inter-component interaction can be tuned by Feshbach resonances \cite{Inouye1998a}.

Recently, polarons were observed in an ultracold atomic Fermi gas \cite{Schirotzek2009a, Nascimbene2009a}. In \cite{Schirotzek2009a} 
radio-frequency spectroscopy was used to measure
a sharp quasiparticle peak solely for the minority component in the highly imbalanced gas of $^6$Li atoms.
Theoretically, various approaches such as Monte Carlo studies \cite{Prokof2008a, Prokof2008b}, 
$T$-matrix approaches \cite{Combescot2007a, Punk2007a, Bruun2010a} and variational ans\"atze \cite{Chevy2006a, Combescot2007a, Punk2009a} have 
been used to model the phenomenon. A variational ansatz by F. Chevy \cite{Chevy2006a} has explained the qualitative features of the experiment.
In \cite{Punk2009a}, variational ansatz describing also the BEC side of the Feshbach resonance, including molecular formation, was proposed,
and the ground state properties matched well with the Monte Carlo studies \cite{Prokof2008a}.

In one-dimensional optical lattices, exact methods provide straightforward 
approaches for highly polarized gases \cite{Feiguin2007a, Rizzi2008a, Batrouni2008a, Tezuka2008a, Luscher2008a}. 
In this article, we use an exact numerical method to investigate highly imbalanced Fermi gases
from the polaronic point of view. We study both the ground state properties and the dynamics, and make a comparison to two different
approximative solutions. In the weakly interacting limit the results are compared to the polaron ansatz of \cite{Chevy2006a}, and in the strongly
interacting limit the results are compared to an approximative solution of the Bethe ansatz. 
In both limits, an excellent agreement between the exact and approximative solutions is found.
However, although we analyze numerical results using a polaron ansatz, we do not necessarily claim the 
excistence of a polaron in a one-dimensional system: well-defined polaronic quasiparticles may not exist due to the 
one-dimensional nature of the system. We discuss implications of our results to three-dimensional systems.

In Section \ref{sec:ground_state} we consider ground state properties of highly imbalanced gases. In Section \ref{sec:rf} 
we study radio-frequency response
of the ground states. In Section \ref{sec:discussion} limitations of our model as well as the connection to higher dimensional 
systems is discussed. In Section \ref{sec:conclusions} we present the conclusions. 

\section{Exact simulations}
\label{sec:ground_state}
For atoms in an optical lattice, the physics is well captured by the Fermi-Hubbard model
  \begin{equation}
    \hat{H}_0
    = -J \sum_{\langle i, \, j \rangle \sigma} \hat{c}^\dagger_{i\sigma} \hat{c}_{j\sigma} +
    U\sum_i \hat{n}_{i \uparrow}\hat{n}_{i \downarrow},
  \label{eq:hubbard_hamiltonian}
  \end{equation}
where $J$ is the hopping energy, $U$ is the on-site interaction strength between different spin components 
$\sigma \in \{\uparrow,\downarrow\}$, 
$\hat{c}_{i\sigma}$ ($\hat{c}^\dagger _{i\sigma}$) 
annihilate (create) a fermion for a site $i$ with spin $\sigma$ and $\hat{n}_{i\sigma} =\hat{c}^\dagger _{i\sigma}\hat{c} _{i\sigma}$. 
Here we consider only attractive 
interactions $U < 0$. In recent experiments in \cite{Schirotzek2009a, Nascimbene2009a} in highly polarized gases qualitative features have been explained with the variational ansatz \cite{Chevy2006a}
  \begin{equation}
  \begin{split}
    |\Psi \rangle &= \varphi _0 \, \hat{c}^\dagger _{0 \downarrow} \, |FS\rangle _\uparrow \, |\emptyset \rangle _\downarrow \\ &+ 
    \sum_{q < k^\uparrow _F, k> k^\uparrow _F} \varphi _{kq} \, \hat{c}^\dagger _{k \uparrow} \, \hat{c}_{q\uparrow} \, 
    \hat{c}^\dagger _{q - k \downarrow} \,
    |FS\rangle _\uparrow \, |\emptyset \rangle _\downarrow,
  \label{eq:ansatz_chevy}
  \end{split} 
  \end{equation}
where $|FS\rangle$ refers to a filled Fermi sea, $|\emptyset\rangle$ to the vacuum state and $k^\uparrow _F$ is a Fermi momentum for spin 
$\uparrow$ particles. 
In the first term the minority particle lies at the bottom of the band and majority atoms form the Fermi sea. 
In the second term the minority atom is scattered out from the lowest momentum state and the filled Fermi sea
of majority particles is broken to particle--hole excitations. The variational coefficients $\varphi _0$ and $\varphi _{kq}$ are found 
by minimizing the energy. The energy difference between the interacting and non-interacting ground states is often named
as the polaron energy $E_p = E_g - E^{non-int}_g$ because this energy difference corresponds to the energy needed to add a single impurity. 
Note that with our definition $E_p < 0$ for attractive interaction strengths.
The quantity $Z = |\varphi _0|^2$ is a measure of the quasiparticle weight of the polaron.
  \begin{figure}[t]
    \includegraphics[width = 0.45\textwidth]{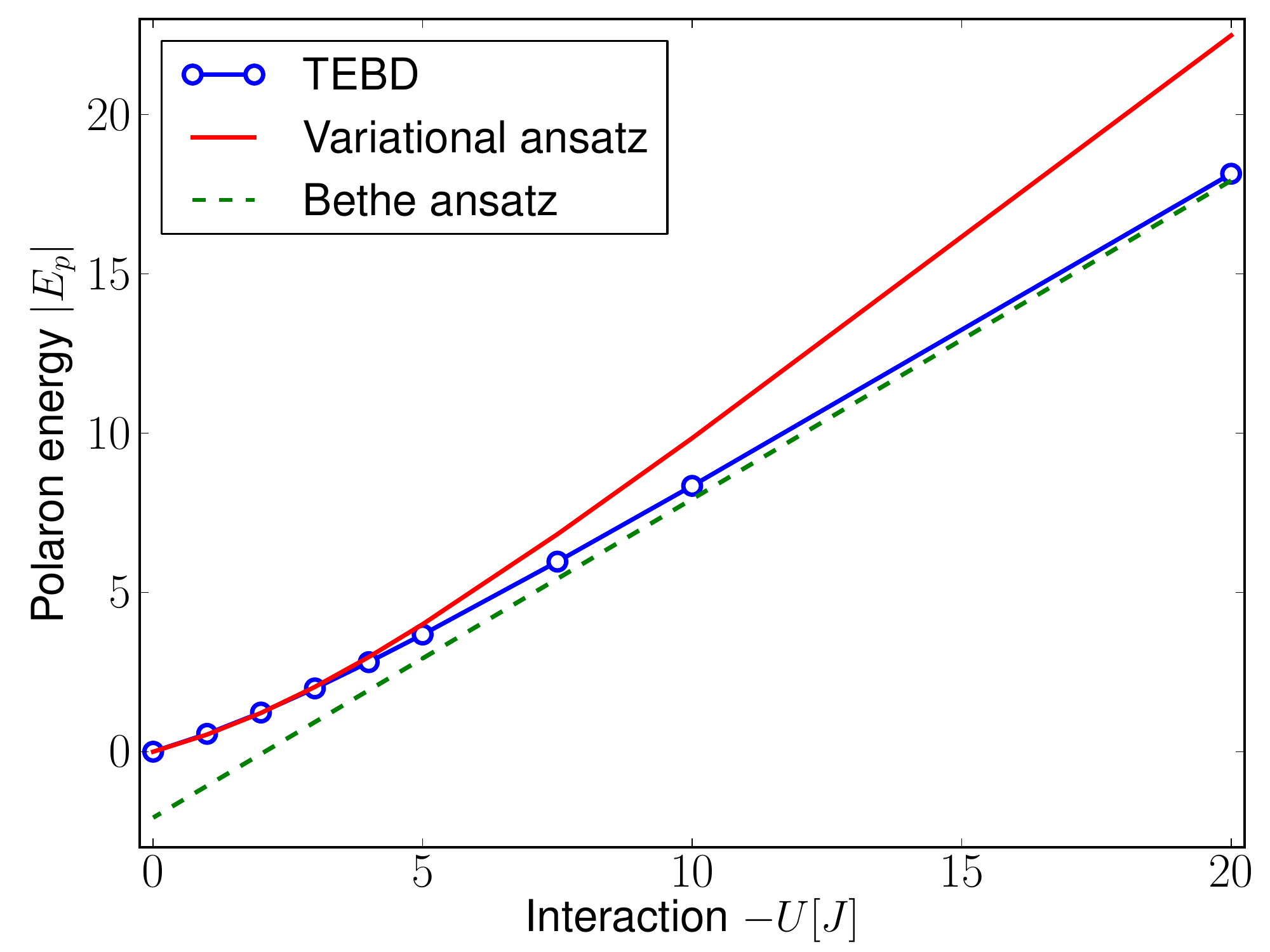}
    \caption{The polaron energy $E_p$ as a function of the interaction strength ($-U$). The weakly interacting regime 
    is well described by the ansatz of Eq. \eqref{eq:ansatz_chevy} (red solid line) and the strongly interacting limit 
    with the Bethe ansatz with $-U/J \rightarrow \infty$ (green dashed line). 
    The number of sites is $N_L = 40$ and the atom numbers are $N_\uparrow = 20$, $N_\downarrow = 1$.}
    \label{fig:energy_polaron}
  \end{figure}
  \begin{figure}
    \includegraphics[width = 0.45\textwidth]{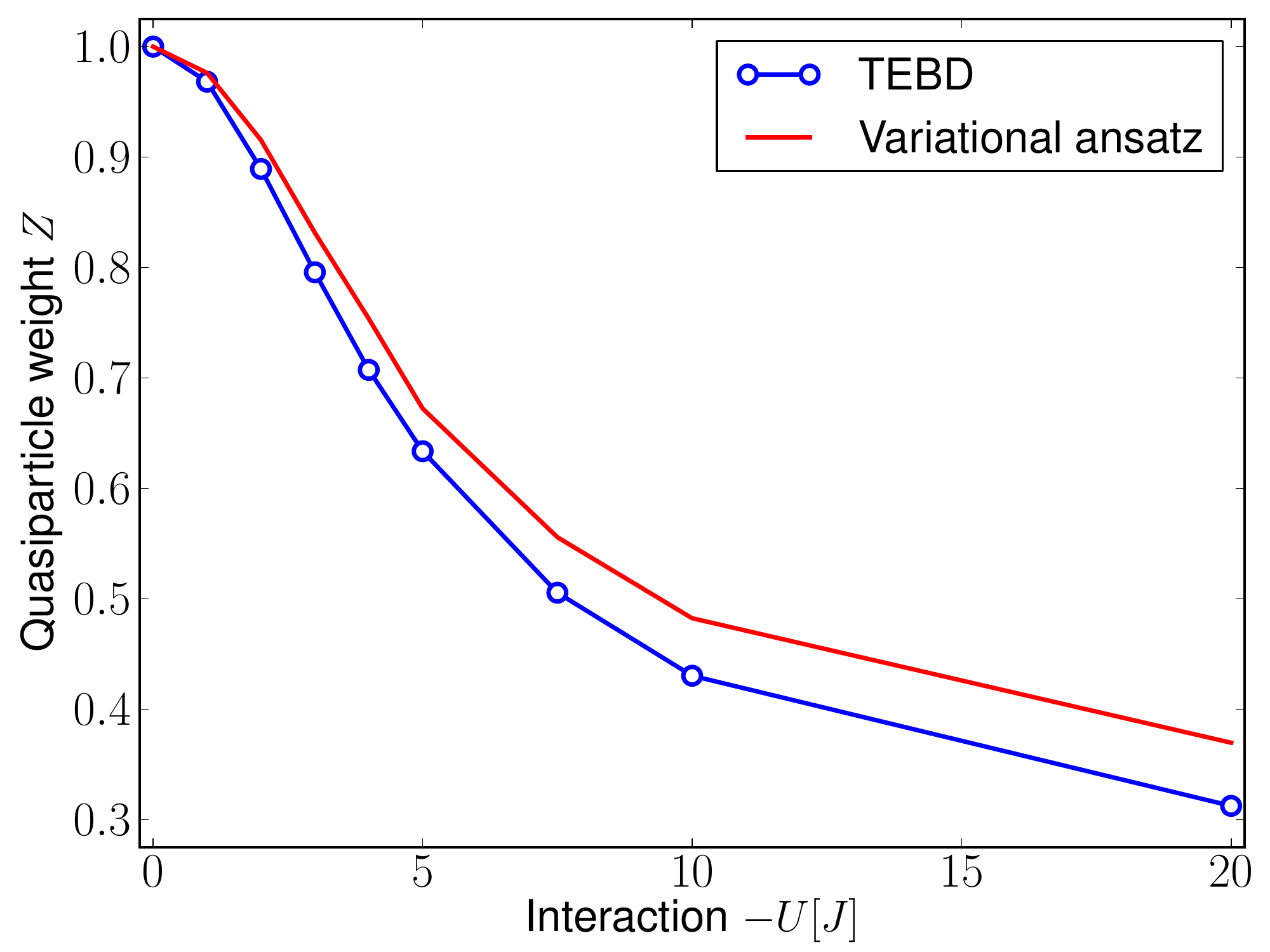}
    \caption{The quasiparticle weight $Z = |\varphi _0 |^2$ as a function of the interaction strength $-U$. Parameters are as in Fig.
    \ref{fig:energy_polaron}.}
    \label{fig:overlap}
  \end{figure}

In order to see how well the ansatz describes the system we have calculated the exact ground state using 
the time-evolving block decimation (TEBD) algorithm \cite{Vidal2003a} which allows us to determine several observables.
The polaron energy is the expectation value $E_p = \langle g| \hat{H}_0 | g \rangle - E^{non-int}_g$ and the quasiparticle weight is the square of the innerproduct 
$Z = |\langle g \,|\, g^{non-int} \rangle |^2$ where $|g\rangle$ and $|g^{non-int}\rangle$ are ground states for the interaction strengths 
$U < 0$ and $U = 0$, respectively, both 
with same particle numbers. In Fig. \ref{fig:energy_polaron} the polaron energy $E_p$ and in Fig. \ref{fig:overlap} the quasiparticle weight for the ansatz and exact numerics
are shown. For small interaction strengths $U\lesssim 5J$ the results are in good agreement but start to deviate for strong attractions. 

In three-dimensional free space the breakdown of the ansatz \eqref{eq:ansatz_chevy} in the BEC limit has been shown experimentally 
\cite{Schirotzek2009a} and described theoretically \cite{Punk2009a}. In the limit $1/(k_F a) \ll 1$ the emergence of molecular states is expected. However, in a one-dimensional optical lattice molecules are not formed even for strong, attractive interactions $-U \gg 1$. 

\begin{figure}[t]
    \includegraphics[width = 0.45\textwidth]{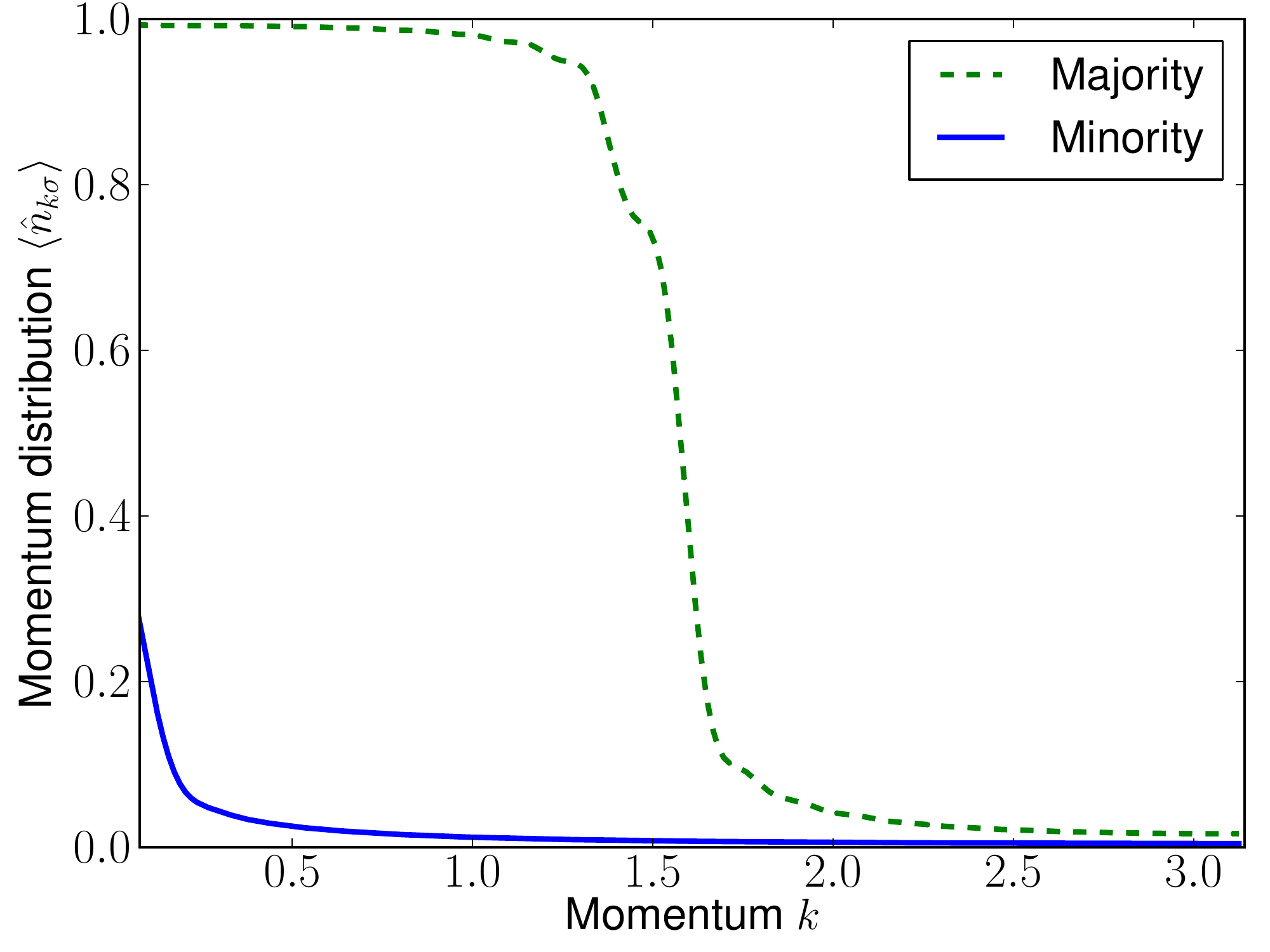}
    \caption{Momentum distributions $\langle \hat{n}_{k, \sigma} \rangle = \langle \hat{c}^\dagger _{k \sigma} \hat{c}_{k\sigma}\rangle$ 
    for the interaction strength $U = -10$. The number of lattice sites is $N_L = 40$ and 
    the atom numbers are $N_\uparrow = 20$, $N_\downarrow = 1$. The pairing emerges between atoms close to the 
    two Fermi surfaces, no pairing between $k$ and $-k$ above $k^\uparrow _F$ is present.}
    \label{fig:momentum1}
  \end{figure}

M. Punk et al. \cite{Punk2009a} provided an ansatz for the molecular regime. In their ansatz the dominant contribution comes from the term
  \begin{equation}
    \sum_{k>k_F^\uparrow} \phi _{k} c^\dagger _{\downarrow -k} c^\dagger _{\uparrow k} c_{\uparrow 0} 
    |FS\rangle _\uparrow \, |\emptyset \rangle _\downarrow,
  \end{equation}
where $\phi _{k}$ are variational coefficients. 
The summation is now restricted to momenta above the Fermi momentum $k^\uparrow _F$ of majority particles. The momentum distributions 
for minority and majority components in our case are shown in Fig. \ref{fig:momentum1}. The momentum distribution for minority component
is extremely small above the Fermi momentum $k^\uparrow _F$, and therefore the ansatz for molecular regime
in highly polarized gases does not improve the ansatz \eqref{eq:ansatz_chevy} in one-dimensional lattices.
However we are able to explain our results
in the strongly interacting limit quantitatively with the Bethe ansatz (BA). In the limit $-U/J \rightarrow \infty$ the 
Bethe ansatz solution can be approximatively mapped to spinless fermions \cite{Massel2009a}. The energy becomes (see Appendix)
  \begin{equation}
  \begin{split}
  E =& \, U - 2J \sum_{i=1}^{N_\uparrow - N_\downarrow} \text{cos}(k_j) \\
  k_j =& \frac{\pi j}{N_L + 1}. %, \hspace{0.5cm} j \in \{0, 1, ..., N' \}
  \label{eq:bethe_ansatz}
  \end{split}
  \end{equation}
For $N_\uparrow = 20$, $N_\downarrow = 1$ and $N_L = 40$ the polaron energy results from the calculation
  \begin{equation}
  \begin{split}
    E_p =& \, E(U, N_\uparrow, N_\downarrow) - E^{non-int}(U = 0, N_\uparrow, N_\downarrow) \\ 
    =& \, U -2J\sum_{j = 1}^{19}\text{cos}(k_j) \\ &\, - \left(- 2J\sum_{j = 1}^{20}\text{cos}(k_j) - 2J\text{cos}(k_1) \right) \\
    =& \, U +  2J\text{cos}(k_1) + 2J\text{cos}(k_{20}) \approx U + 2.07J, %1.98J.
  \label{eq:bethe_ansatz_polaron}
  \end{split}
  \end{equation}
where $E(U, N_\uparrow, N_\downarrow)$ ($E^{non-int}(U = 0, N_\uparrow, N_\downarrow)$) is the energy for the system with 
interaction strength $U$ ($U = 0$) and particle numbers $N_\uparrow, N_\downarrow$ ($N_\uparrow, N_\downarrow$).
The BA result is shown in Fig. \ref{fig:energy_polaron} with the exact result. In the strongly interacting limit $-U \gtrsim 7J$ 
the Bethe ansatz solution is in good agreement with the exact solution.

\section{Radio-frequency spectroscopy}
\label{sec:rf}
In the experiment \cite{Schirotzek2009a}, radio-frequency (rf-) spectroscopy was performed to the minority and majority spin components. In 
rf-spectroscopy one of the spin components, $\uparrow$ or $\downarrow$, is coupled to a third spin state, denoted by 3, which is 
not populated and sufficiently weakly 
interacting with the initial states. Theoretically, the rf-field is well described in the rotating wave approximation 
by the Hamiltonian
  \begin{equation}
    \hat{H}^\sigma _{rf}(t) = \Omega \sum_j ( e^{-i\delta t} \hat{c}^\dagger _{j \sigma} \hat{c} _{j3} + 
    e^{i\delta t} \hat{c}^\dagger _{j 3} \hat{c} _{j\sigma}),
  \end{equation}
where the sum is over the lattice sites, $\Omega$ is the coupling strength and $\delta$ is the detuning of the rf-field 
from the $\sigma - 3$ transfer frequency.

From the variational ansatz \eqref{eq:ansatz_chevy} we have straight access to the rf-spectra through the Fermi golden rule
  \begin{equation}
  \begin{split}
    I_\downarrow &\propto 
    \sum_f |\langle f | \hat{H}^\downarrow _{rf} | g \rangle|^2 \, \delta ^{(1d)}(\delta - E_f + E_g) \\ 
    &= |\varphi _0|^2 \, \delta ^{(1d)} (\delta - |E_p|) + \Gamma^{inc}(\varphi _{k q}, \delta),
  \label{eq:variational_spectrum}
  \end{split}
  \end{equation}
where the summation is over all states with energies $E_f$, $|g\rangle$ is the ground state and $E_g$ its energy \cite{Schirotzek2009a}.
The first term gives rise to a narrow peak at the polaron energy and behind that is a broad tail resulting from the term 
$\Gamma^{inc}(\varphi _{k q}, \delta)$ describing the incoherent part. 
Qualitatively the measured spectra \cite{Schirotzek2009a} matched with the variational ansatz for the 
attractive interaction strenghts and close to the unitary limit. However, in the BEC side $1/(k_F a) \ll 1$ the minority 
and the majority spectra overlapped completely which signaled molecular pairing.
  \begin{figure}[t]
    \includegraphics[width = 0.45\textwidth]{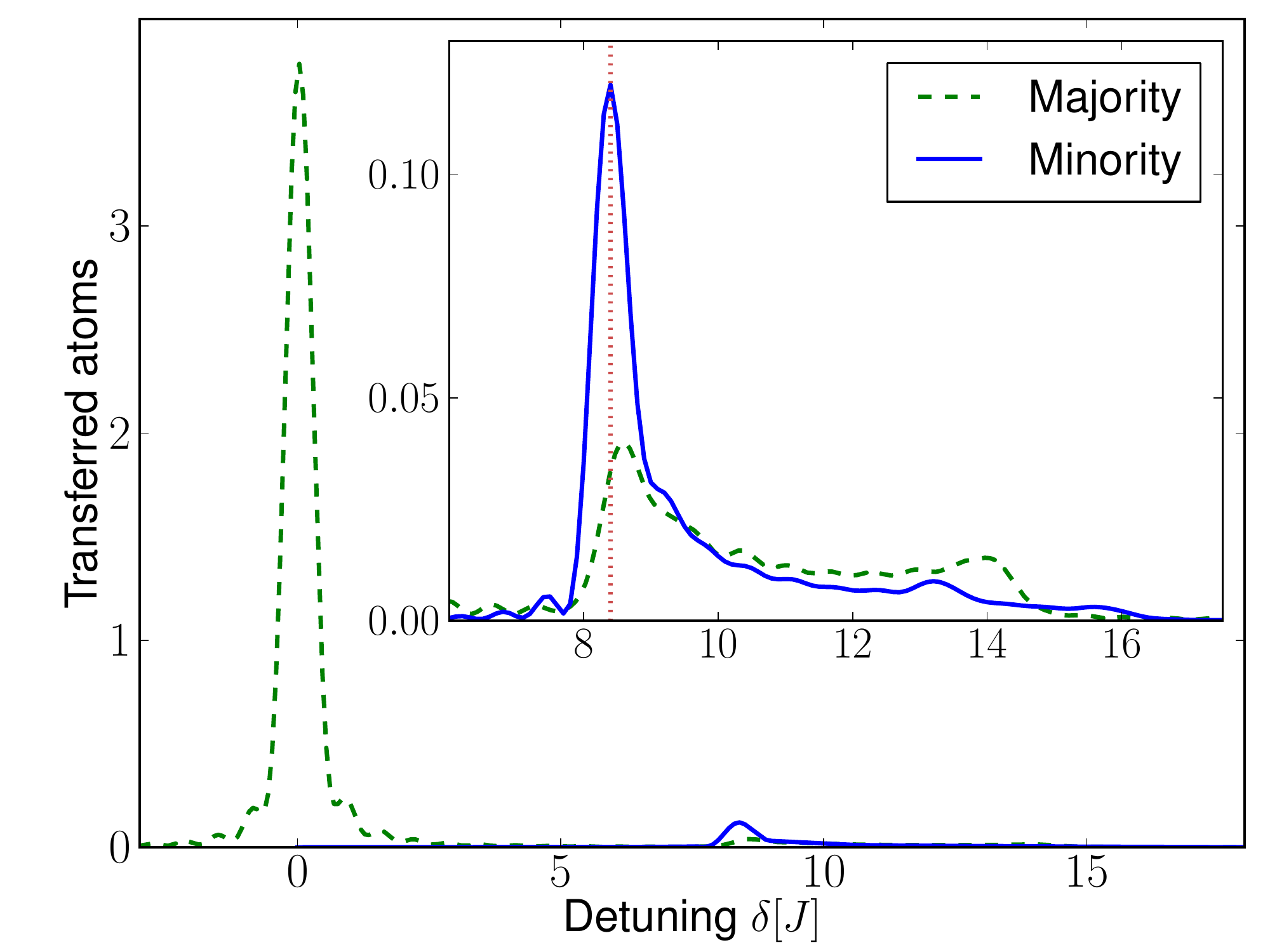}
    \caption{Minority and majority spectrum for $U = -10$. The inset is a zoom of higher detunings where a quasiparticle peak emerges. 
             The vertical line is the polaron energy $E_p$ calculated from the exact ground state. 
             The number of spin components are $N_\uparrow = 20$, $N_\downarrow = 1$ and 
             the number of the lattice sites is $N_L = 40$. Spectra for other interactions are qualitatively similar.}
    \label{fig:spektri_u-10}
  \end{figure}
The TEBD-algorithm allows us to calculate the full time-evolution of the system, and therefore we can evaluate the relation between 
the rf-spectra and the polaron-like state. To obtain the spectra, we first calculate the ground state of the Fermi-Hubbard Hamiltonian 
\eqref{eq:hubbard_hamiltonian} and then the dynamics by operating on the ground state with the time-evolution operator 
$\exp (-i \hat{H} t)$ consisting of the Hubbard Hamiltonian and the rf-field:
  \begin{equation}
    \hat{H} = \hat{H}_0 + \hat{H}^\sigma _{rf}. 
  \end{equation}
The spectra for the minority and 
majority components calculated using exact numerics are shown in Fig. \ref{fig:spektri_u-10}. In the majority spectrum the main contribution is 
at zero detuning because of the unpaired atoms, but some atoms are transferred also when $\delta \sim 8 - 14J$. 
The minority spectra is highly peaked at $\delta = 8.4J$, and the peak position is exactly at the polaron energy $E_p$ which
is calculated from the exact ground state, and is very well approximated by Eq. \eqref{eq:bethe_ansatz_polaron}.
At larger detunings a long tail emerges but is cut due to the restriction of momenta in an optical lattice to $k\in [-\pi,\pi]$.

In Eq. \eqref{eq:variational_spectrum} the particle--hole excitations give rise to the incoherent part of the spectra
  \begin{equation}
    \Gamma^{inc}(\varphi _{k q}, \delta) = |\varphi _{kq}|^2 \, \delta^{(1d)}
    ( \delta - \epsilon _{q-k} - \epsilon _k + \epsilon _q + \epsilon _0 - |E_p|),
  \end{equation}
where $\epsilon _k = -2J\text{cos}(k)$ is the dispersion relation for non-interacting particles in a lattice.
The minimum detuning which contributes to the minority spectrum is for $q = k$ and gives 
$\delta _{min} = E_p$. The maximum detuning arises when the system has the hole in the bottom of the Fermi sea $q = 0$ and the excitation 
lies at the van Hove singularity $p = \pi$ \emph{i.e.}
$\delta _{max} =  \epsilon _{-\pi} + \epsilon _\pi - \epsilon _0 + E_p - \epsilon _0 = E_p + 8J$. Now, the width of the spectrum is 
$\delta _{max} - \delta _{min} = 8J$. This is in good agreement with the exact spectra.

  \begin{figure}[t]
    \includegraphics[width = 0.45\textwidth]{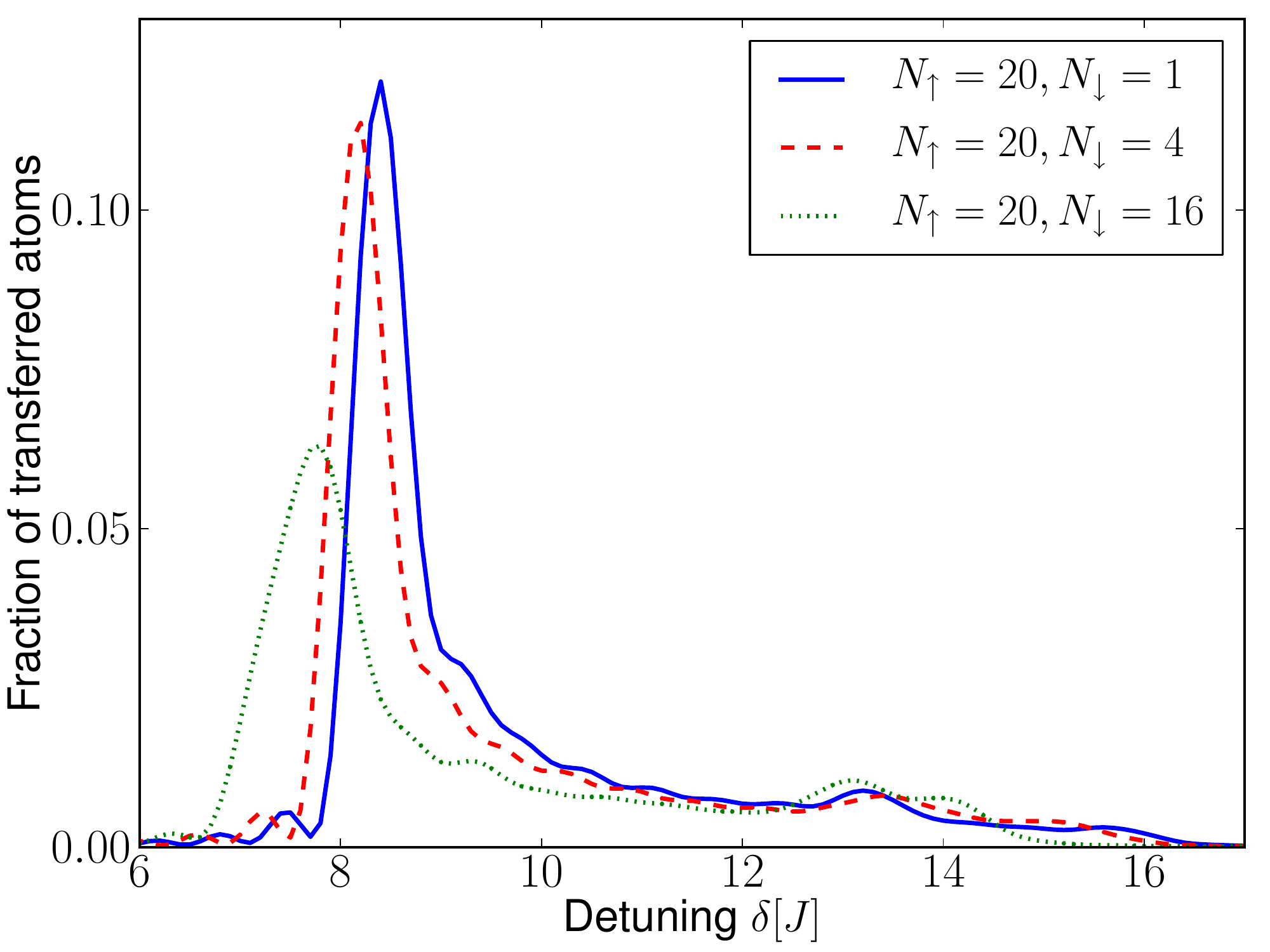}
    \caption{Minority component spectra for various imbalances. The interaction strength is $U = -10J$.}
    \label{fig:spectra_minority_polarizations}
  \end{figure}
In Fig. \ref{fig:spectra_minority_polarizations} we have varied the polarization. 
The spectra look similar for large spin imbalance $N_\downarrow/N_\uparrow \lesssim 0.2$ but for small
spin imbalance the spectra get broadened and are shifted to lower detunings. Let us analyze the peak positions with the Bethe ansatz.
The Fermi golden rule states that the peak position comes from the energy difference between the ground state and 
final state. For ground state with $N_\uparrow = 20$ and $N_\downarrow = 2$ the final state for the Fermi golden rule analysis 
has the particle numbers $N_\uparrow =20$, $N_\downarrow = 1$ and $N_3 = 1$. Therefore, the energy difference becomes
  \begin{equation}
  \begin{split}
    \Delta E =& \, E_G - E_F \\
    =& \, 2U - 2J \sum_{i=1}^{18} \text{cos}(k_i) - \\ 
    & \, \left(U - 2J\sum_{i=1}^{19} \text{cos}(k_i) - 2J\text{cos}(k_{final}) \right) \\
    =& \, U + 2J\text{cos}(k_{19}) + 2J\text{cos}(k_{final}). 
  \end{split}
  \end{equation}
The lowest contribution to the spectrum arises when $k_{final} = k_1$ which implies
  \begin{equation}
    |E_p| - |\Delta E| = 2J\text{cos}(k_{19}) - 2J\text{cos}(k_{20}) \approx 0.15.
  \end{equation}
The energy difference $|\Delta E|$ is smaller than the polaron energy $|E_p|$ (energy difference in the case of a single minority component), 
and therefore the spectrum 
shifts to lower detunings when polarization is increased. Furthermore, we can make quantitative comparison of this Bethe ansatz result
to the exact numerics. The distance between the two peaks in the rf-spectrum for the 
cases $N_\downarrow = 1, 2$ is around $0.1$, which is in good agreement with the Bethe ansatz result $|E_p| - |\Delta E| = 0.15$.

\section{Discussion}
\label{sec:discussion}
We have shown that in a highly
imbalanced Fermi gas, the ground state exhibits polaronic-like
behaviour. The quasiparticle nature of this excitation, which can be
deduced from the non-zero quasiparticle weight of polaron, is
confirmed by the rf-spectroscopy analysis. Two considerations are,
however, in order. The first concerns the dynamical properties of the
quasiparticle.  From our analysis it is not possible to prove the
stability of the polaronic-like particle propagation through the cloud
of majority atoms. The polaron weight might split up into
particle-hole excitations. The rf-spectroscopy, measuring the single
particle spectral function, does not describe collective
properties. A nonzero value for quasiparticle residue would correspond
to dynamical stability of the quasiparticle: 
\begin{equation}
\lim_{t\rightarrow\infty} |G_\downarrow(\mathbf{k}=0, t)| \neq 0.
\end{equation} 
The second consideration relates to the quasiparticle
description of the polaron. On general grounds it is well established
that one-dimensional systems exhibit a "collectivization" of the
excitation, the typical example being the spin-charge separation in
the \emph{balanced} Fermi-Hubbard Hamiltonian. However it has been
shown that in the case of an \emph{imbalanced} gas the spin-charge
separation is violated \cite{Feiguin2007a, Luscher2008a, Tezuka2008a,
Rizzi2008a}.  We expect that the validity of the ansatz given in Eq.
\eqref{eq:ansatz_chevy}, and hence the validity of the polaron
quasiparticle description even in 1D case, in the limit of weak
interaction, can be pictured as the extreme limit of the violation of
the spin-charge separation.

\section{Conclusions}
\label{sec:conclusions}

We have investigated the ground state properties and the rf-spectrum
of a highly imbalanced Fermi gas in a one-dimensional optical
lattice. The exact numerical results can be explained in terms of the
variational ansatz given in Eq. \eqref{eq:ansatz_chevy} and of Bethe
ansatz equation (in the limit of $-U/J\rightarrow \infty$) for 
$-U \lesssim 5J$ and $-U \gtrsim 7J$, respectively. Our results suggest the possibility of
the existence of a polaronic quasiparticle, further analysis is
however required to investigate its dynamical properties. Moreover, we
would like to point out that the setup proposed here is well within
reach of the current experimental capabilities.
Our analysis provides exact numerical results for comparison with future experiments, 
as well as effective physical interpretation for both the weak and strong interaction limits.
\\
{\it Acknowledgements}
We thank J.J. Kinnunen for useful discussions. This work was supported by the National Graduate School 
in Materials Physics and Academy of Finland (Project No. 213362, No. 217045, No. 217041, No. 217043), 
and conducted as a part of a EURYI scheme grant, see www.esf.org/euryi. We acknoledge the use of
CSC -- IT Center for Science Ltd computing resources in this work. Correspondence should be addressed
to P\"aivi T\"orm\"a (paivi.torma@hut.fi).

\section*{Appendix}
Eq. \eqref{eq:bethe_ansatz} can be obtained considering the BA solution for the
open-boundary conditions (OBC) Fermi-Hubbard model in the limit $U/J
\to \infty$.  Analogously to the calculations performed in \cite{Massel2009a}, it is possible to prove that the excitations of the
system, for a repulsive interaction, can be described in terms of
$N=N_\uparrow+N_\downarrow$ spinless fermions with energy and momenta
given respectively by
\begin{align}
  E&=-2J\sum_{j=1}^{N}\cos k_{j} \nonumber\\
  k_{j}&=\frac{\pi}{L+1}I_{j} \quad I_{j}\in\mathbf{N} \textrm{,}\, j=\left[1\ldots N\right]. 
  \label{eq:e_k}
\end{align}
The expression for $k_j$ in Eq. \eqref{eq:e_k} is derived from the BA
equations for \cite{Asakawa1996a}.  $N_{\uparrow}$ up, $N_{\downarrow}$ down
electrons on $N_L$ sites BA equations can be written as
  \begin{equation}
  \begin{split}
    \label{eq:BAobc1}
    2L k_j=& 2\pi I_j-2 k_j-
          \sum_{\beta=1}^{N_{\uparrow}} 
            \bigg[
                \Phi\left(2\frac{\sin(k_j)-\lambda_\beta}{u}\right) \\
             & + \,   \Phi\left(2\frac{\sin(k_j)+\lambda_\beta}{u}\right)                 
            \bigg]
  \end{split}
  \end{equation}
  and
  \begin{equation}
  \begin{split}
    \label{eq:BAobc2}
      & \sum_{j=1}^{N_{\uparrow}+N_{\downarrow}}
            \bigg[
                \Phi\left(2\frac{\lambda_\alpha-\sin(k_j)}{u}\right) + 
                \Phi\left(2\frac{\lambda_\alpha+\sin(k_j)}{u}\right)                 
            \bigg] \\ =& 
         2\pi J_\alpha+\sum_{\beta=1  (\beta \neq \alpha)}
    \bigg[
                \Phi\left(\frac{\lambda_\alpha-\lambda_\beta}{u}\right) +
                \Phi\left(\frac{\lambda_\alpha+\lambda_\beta}{u}\right)                 
            \bigg],
  \end{split}
  \end{equation}
  where $j=1,\dots, N_{\uparrow}+N_{\downarrow}$, $\alpha=1,\dots,N_{\downarrow}$,
  $I_j\,,\,J_\alpha \in \mathbb{N}$, $U/J=u$,
  $\Phi(x)=2\,\tan^{-1}(2\,x)$ and $\lambda_{\alpha}$ are the spin velocities.

The distribution of $I_{j}$ should correspond to a condition where
the energy is minimized. For a balanced gas at half filling, the energy
minimization condition is given by $I_{j}=\left[1\ldots L\right]$,
leading to $E=-2J\sum_{j=1}^{L}\cos k_{j}=0$, $p=\sum_{j=1}^{L}k_{j}$.

If we take into account the mapping $U\to -U$ and that, in the limit
$U/J\to \infty$, the total number of pairs is equal to
$N_{\downarrow}$, the single-site basis states can be mapped according
to the following scheme
\begin{equation}
  \label{eq:map}
  |\uparrow \downarrow \rangle \leftrightarrow  |\downarrow \rangle, \quad
  |\emptyset \rangle \leftrightarrow  |\uparrow \rangle, 
\end{equation} 
leading to $N=N_L-(N_{\downarrow}-N_{\uparrow})$, and hence 
\begin{equation}
    \label{eq:en_1}
     E=-2J\sum_{j=1}^{N_L-(N_{\downarrow}-N_{\uparrow})}\cos k_{j}
\end{equation}
which, taking into account the fact that
$\sum_{j=1}^{N_L}\cos k_{j}=0$, can be  written as
\begin{equation}
  \label{eq:en_2}
   E=-2J\sum_{j=1}^{N_{\uparrow}-N_{\downarrow}}\cos k_{j}.
\end{equation}
\newpage
\bibliographystyle{apsrev}
\bibliography{paperi.bib}

\end{document}